\documentclass[iop]{emulateapj}

\slugcomment{Accepted for publication in the ApJ Letters}

\shorttitle{Super-Li rich turn-off star in NGC~6397}
\shortauthors{A. Koch et al.}

\begin{document}

\title{Discovery of a super-Li rich turn-off star in the metal poor globular cluster NGC~6397\altaffilmark{$\dagger$}}	

\author{Andreas Koch\altaffilmark{1},  
Karin Lind\altaffilmark{2}, 
\& R.Michael Rich\altaffilmark{3}}
\altaffiltext{$\dagger$}{This paper includes data gathered with the 6.5 meter Magellan Telescopes located at 
Las Campanas Observatory, Chile.}
\altaffiltext{1}{ZAH, Landessternwarte K\"onigstuhl, Heidelberg, Germany}
\altaffiltext{2}{Max-Planck-Institut f\"ur Astrophysik, Garching, Germany }
\altaffiltext{3}{University of California Los Angeles, Department of Physics \& Astronomy, Los Angeles, CA, USA}
\email{akoch@lsw.uni-heidelberg.de}
\begin{abstract}
We report on the discovery of a super-Li rich turn-off star in the old (12 Gyr), metal poor ([Fe/H]=$-2.1$ dex)  globular cluster (GC) NGC~6397, based on high-resolution MIKE/Magellan spectra.  This star shows an unusually high lithium abundance of A(Li)$_{\rm NLTE} = 4.03\pm0.06\pm0.14$ dex (or, 4.21, accounting for possible contamination from a binary companion) that lies above the canonical Li-plateau by a factor of 100. This is the highest Li enhancement found in a Galactic GC dwarf star to date. 
We discuss several enhancement mechanisms, but none can unambiguously explain such a high overabundance. 
The spectrum of the star shows a possible indication of binarity, but its line strengths and chemical element abundance ratios   are fully compatible with other turn-off stars in this GC, seemingly ruling out mass transfer from an AGB companion as origin of the high A(Li). 
A possible cause is an interaction with a red giant that has undergone cool bottom processing. 
\end{abstract}
\keywords{Stars: abundances --- stars: binaries ---  stars: Population II --- globular clusters: individual (NGC~6397) --- nuclear reactions, nucleosynthesis, abundances}
\section{Introduction}
\setcounter{footnote}{4}
Lithium remains as one of the crucial chemical elements that ties cosmology to important aspects of stellar structure and evolution.
While Big Bang Nucleosynthesis (BBN) established a primordial abundance\footnote{A($X$)$\equiv$
12+$\log$($N_X / N_H$).} of A(Li)$=$2.72 dex (Cyburt et al. 2008), 
values found in Galactic halo dwarfs are collectively lower (Spite \& Spite 1982), at 2--2.4 dex, and even larger depletions are seen for main sequence stars below $\sim$5800 K. 
Thus the primordial abundance must have been uniformly depleted by a factor of $>$3, indicating the impact of intrinsic processes 
that allow for the diffusion or mixing of Li into the hotter stellar regions where it is easily destroyed by  $p$-, and $\alpha$-reactions at a few $\times 10^6$ K.

To further constrain such intrinsic processes, it is important to investigate what mechanisms can modify the stellar Li content, ideally by probing its dependency on evolutionary status (i.e., mass, $M_V$, or T$_{\rm eff}$). 
As the latter are easily determinable in star clusters of given age and distance, these pose the ideal objects for studying  such internal stellar processes. 
Metal poor globular clusters (GCs) are particularly attractive, since they trace the early conditions of the Universe. 
Moreover, the occurrence of light element variations and multiple stellar populations in GCs (e.g., Gratton et al. 2004) may provide important clues as to the origin of the Li anomalies.

{\em Enhanced} Li values, on the other hand, are more difficult to explain
and efficient Li{\em-production} mechanisms need to be invoked.  
Specifically, the occurrence of stars with an A(Li) higher than even the ISM abundance argues against a simple preservation of the primordial value in the stars.  
Amongst the possible origins of strong Li-overabundances 
are: the concentration of Li in spots of magnetic stars (e.g., Kochukhov 2008), production in cosmic rays (Reeves 1970), 
accretion of substellar bodies (e.g., Takeda et al. 2001), 
joint actions of thermal instabilities of a Li-burning shell and rotation-induced mixing (Palacios et al. 2001),  
or the pollution by luminous (post-) asymptotic giant branch (AGB) stars of a  certain mass range (Abia et al. 1993; Sackman \& Boothroyd 1999). 

The latter enhancements are obtained via a $^7$Be-transport mechanism (Cameron \& Fowler 1971; hereafter CF71), which acts at temperatures of a few $\times 10^7$ K and needs to instigate the 
$^3$He($\alpha$, $\gamma$)$^7$Be and $^7$Be(e$^-$, $\nu$)$^7$Li reactions via circulations between the stellar interior and the convective envelope.  
In low-mass giants, however, regions with the required high temperatures are not in contact with the surface and extra-mixing is needed, 
which is usually identified with cool bottom processing (CBP;  Wasserburg et al. 1999; Sackman \& Boothroyd 1999; see also Smith et al. 1999; Monaco et al. 2011).

Here we report the discovery of a turn-off (TO) star in the metal poor Galactic GC NGC~6397 
that shows an extraordinary Li-overabundance. 
While there exists a smattering of (super-) Li rich {\em giants} in GCs (e.g., Boesgaard et al. 1998; Smith et al. 1999; Kraft et al. 1999)  and the field (e.g., Brown et al. 1989;  Monaco et al. 2011), Li-rich TO stars that lie above the main-sequence maximum value of A(Li)=3.3 dex (e.g., Lambert \& Reddy 2004) are 
rare, and the record holder to date, at 4.29 dex\footnote{Note, however, that this value is not corrected for NLTE, with expected corrections of $\sim - 0.2$ dex (Lind et al. 2009a).}, is a member of a young (700 Myr), open cluster  (Deliyannis et al. 2002). 
\section{Data}
In Koch \& McWilliam (2011; hereafter KM11) we performed a differential abundance analysis of red giants and TO stars in NGC~6397 
based on high-resolution spectroscopy obtained with the Magellan Inamori Kyocera Echelle (MIKE) instrument at the 6.5-m Magellan2/Clay Telescope. These data reach a resolution of $R\sim40000$ (see KM11 for details of the observation and analysis). 

Our initial sample contained a target with the same color and magnitude as the two TO stars published in KM11 (Table~1), but since a cursory inspection of the spectrum revealed it to be in a binary it was ignored in the previous abundance work. However, this object, \#1657\footnote{ID according to the catalogue of Kaluzny (1997).}, shows unusually strong Li absorption lines at 6103 and 6707\AA~(Fig.~1) so that we felt it warranted to give this spectrum further scrutiny. 
\begin{figure}[htb]
\begin{center}
\includegraphics[angle=0,width=1\hsize]{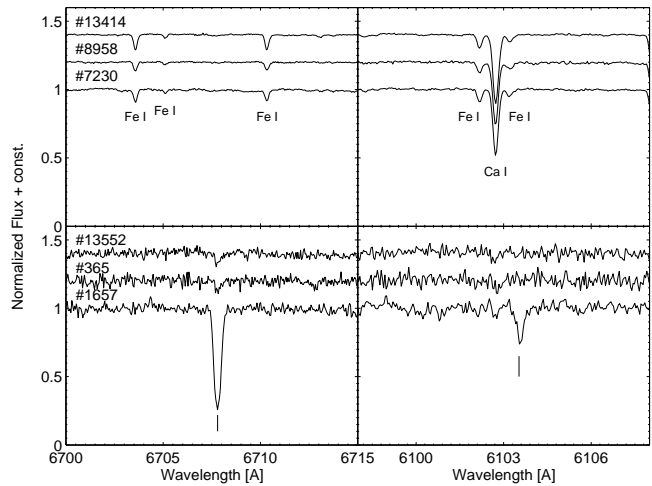}
\end{center}
\caption{Regions around the \ion{Li}{1} 6707 and 6103\AA~lines in the three red giants (top panels) and three TO stars (bottom) from KM11. Spectra have been shifted vertically for clarity.}
\end{figure}
\begin{center}
\begin{deluxetable}{cc}
\tabletypesize{\scriptsize}
\tablecaption{Observations and properties of TO \#1657}
\tablewidth{0pt}
\tablehead{ \colhead{Parameter} & \colhead{Value}}
\startdata
$\alpha$, $\delta$ (J2000.0) &   \phantom{$-$}17:41:06.6, $-$53:43:30.4  \\
V, B$-$V, V$-$K [mag]            & 16.27, 0.57, 1.59 \\
v$_{\rm HC}$ [km\,s$^{-1}$] & 24.8 \\
Date of obs.       & 2005 May 31, Jun 01 \\
Exposure time [s]  & 15000 \\
S/N at 6500\AA     & 50
\enddata
\end{deluxetable}
\end{center}
\subsection{Radial velocities}
To get a clear velocity signal, we cross correlated the spectrum of 
\#1657 against  the presumed single star  \#13552. 
The correlations were carried out on all echelle orders, 
but ignoring  telluric absorption and strong features such as the hydrogen Balmer, \ion{Na}{1} D$_1$ and D$_2$, and \ion{Ca}{2} H and K lines. 

The correlation function in Fig.~2 reveal a strong secondary peak (with 20\% of the main peak's flux signal),  corresponding to a heliocentric radial velocity of $-$95 km\,s$^{-1}$, which is already visible as faint  absorption features of the strongest lines. 
The large velocity separation  of $-120$ km\,s$^{-1}$ to the primary 
renders it unlikely that any of the main star's absorption lines of interest will be contaminated from this object. 
This component is most likely a chance alignment with a faint foreground star and we did not pursue those features any further. 
\begin{figure}[htb]
\begin{center}
\includegraphics[angle=0,width=1\hsize]{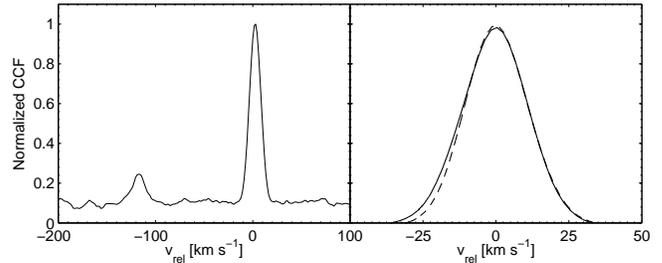}
\end{center}
\caption{Cross correlation functions of the binary TO \#1657 against the single star \#13552 (solid lines) and between two single-lined TOs (dashed line in the right panel).}
\end{figure}

Finally, the correlation signal of the reddest, strongest lines in the spectrum such as the calcium triplet at 8498, 8542, 8662\AA~shows an asymmetry, suggesting a possible binary component with a velocity separation  of 11.5 km\,s$^{-1}$,  although there is only 
marginal indication of radial velocity  variations  across the 21 hours spanned by our five exposures. 
The {\em mean} systemic v$_{\rm HC}$ of 24.8$\pm$1.0 km\,s$^{-1}$ is consistent with cluster membership. 
If present, we estimate that  the potential secondary contributes a mere $\sim$10\%
to the total flux in the slit.

The probability of an overlap with other sources is also verified from the high density on 
archival images from the Hubble Space Telescope's (HST) Advanced Camera for Survey, where  
 at least ten point sources fall within $\sim$$5\arcsec$ around the target (Anderson et al. 2008; Rich et al. 2011). 
While it is evident that our slit orientations picked up additional stars, the HST photometry shows that any such serendipitous 
alignment contributes less than 0.5\% contaminating flux.  
More importantly, further faint objects are likely blended with \#1657: the contours of the actual target star are clearly asymmetric and 
an overlap of sources is highly probable. 
\subsection{Stellar parameters}
In full analogy to KM11 we perform here a differential equivalent width (EW) abundance analysis. 
Given the small separation of the components 
we cannot treat the spectra as 
double-lined and reliably measure individual EWs to determine each component's composition in greater detail (cf. Preston 1994; Thompson et al. 2008). 
In accord with the above, relative flux estimates, the absorption lines in \#1657 are shallower by 
(16$\pm$3)\%  on average compared to the other two, presumed single, TO stars.  
To correct for this continuum veiling,  
 we followed Preston (1994) in diluting all observed  EWs in \#1657 by a factor 1.16.  We will list all our abundance results in Table~2 below based on the observed EWs (``unveiled''), and those multiplied by the veiling factor 1.16 (``veiled''). 

The determination of the stellar parameters is less straightforward than for the other stars in KM11 and leaves room for a broader range of temperatures and in particular microturbulent velocities, $\xi$. This indicates that the contamination from the secondary component is poorly accountable for and predominantly affects the EW plot to accurately derive $\xi$. 
For either set of EWs, a broad range of temperatures 
can reproduce excitation equilibrium and we adopt here a value consistent with the other two TO stars of comparable color in KM11. 
 For the microturbulence we also adopt a value as typically found in the other two TO stars. To first order,  this  yields a flat EW plot, again over a broad range for this parameter. 
We thus adopt  (T$_{\rm eff}$, log\,$g$, $\xi$) = (6282$\pm$250 K, 4.1$\pm$0.2, 1.2$\pm$0.8  km\,s$^{-1}$). 

The resulting mean \ion{Fe}{1} abundance of $-1.93\pm0.06$(stat.)$\pm 0.18$(sys.) dex  is more metal rich than the global GC average from KM11 by 0.17 dex, while 
it is consistent with the TO subset within the uncertainties. 
\subsection{Lithium abundances}
Li abundances for all our targes were obtained from the 6707\AA~resonance line, accounting for its fine structure, and using $gf$-values from Hobbs et al. (1999). Only $^7$Li was used -- the amount of $^6$Li is generally negligible and resulting line asymmetries are below our current spectral sensitivity. 
In practice, we measure EWs of (28, 26) m\AA~for the TO stars (\#13552, 365) and (un)veiled EWs of 325 (377) m\AA~for \#1657. 
No lines could be detected in the red giant branch (RGB) RGB stars and we place upper limits based on a 1$\sigma$ estimate of the local variance (Cayrel 1988). 
All values for A(Li) were confirmed from fitting synthetic spectra to the 6707\AA~line in a least-squares sense, except for the RGB stars with no detectable Li and \#1657, where the line is too strong to be reliably fit with a model profile. In the latter we also detect  the subordinate line at 6103\AA~with an EW of 65 (74) m\AA. 
Finally,  we applied corrections for NLTE effects to our results, based on the extensive parameter grid computed by Lind et al. (2009a). 
While the latter used the MARCS stellar atmospheres (Gustafsson et al. 2008), no systematic difference is found in our use of the Kurucz atmosphere grid  (Castelli \& Kurucz 2004; KM11). 
The NLTE correction for the weak 6103 line in star 1657 is negligible, but, at $-$0.85 dex (for either EW), it has a large impact on the abundance derived from the 6707 \AA~transition. 
It is reassuring that the NLTE results from both lines are in excellent agreement. 

As a result we find an A(Li)$_{\rm NLTE}$ of 2.23 and 2.20 for the two single TO stars, \#13552 and \#365. These values are fully consistent with 
the Li abundances in un-evolved stars (at $\sim$2.0--2.4 dex), as also found in NGC~6397 (Fig.~3; Bonifacio et al. 2002; Gonz\'alez-Hernandez et al. 2009; Lind et 
al. 2009b). 
In fact, their location on the ``Spite-plateau''  (Spite \& Spite 1982) of Pop~II halo dwarfs over a broad range of metallicities 
indicates a uniform depletion mechanism, lowering the primordial abundance left over from BBN via diffusion and mixing processes (e.g., Richard et al. 2005). 
\begin{figure}[htb]
\begin{center}
\includegraphics[angle=0,width=1\hsize]{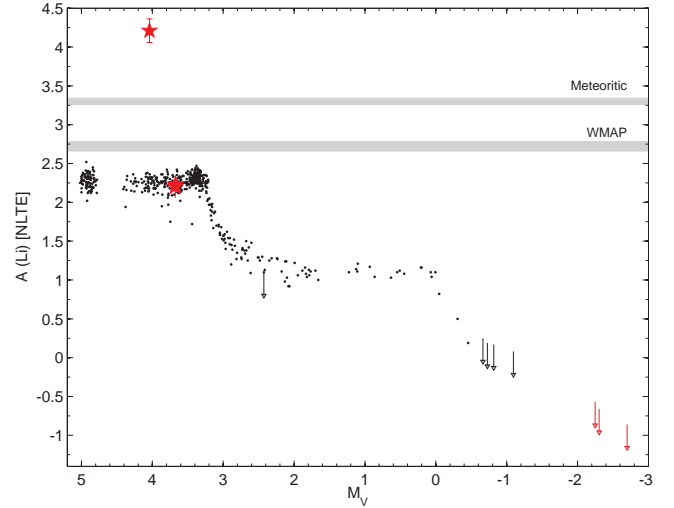}
\end{center}
\caption{Li abundances in NGC~6397 from Lind et al. (2009b; black dots). Our RGB limits are shown as red arrows (at M$_V<-2$), while the TO measurements are highlighted as red stars.}
\end{figure}

For the binary candidate \#1657, however, we find a strongly enhanced value of 
A(Li)$_{\rm NLTE} = 4.21 \pm 0.06$(stat.)$\pm 0.14$(sys.). The quoted errors are based on the 1$\sigma$-scatter of the two lines (``stat.'') and the uncertainties on the input stellar parameters (``sys.'').
The combined, systematic uncertainty is chiefly dominated by
 the uncertainty on T$_{\rm eff}$, while even a large uncertainty on $\xi$ has only a minor contribution of less than 
$\pm0.03$ dex\,(km\,s$^{-1}$)$^{-1}$ to the error budget.
Neglecting the veiling by the overlapping continua would yield an A(Li) lower by 0.18 dex, at a lower [Fe/H] of $-2.04$ dex. In any case, its extraordinarily large EWs place this object undoubtedly in the super-Li rich dwarf star regime.
\subsection{Alpha-elements}
In analogy to KM11 we determine differential abundance ratios of the $\alpha$-elements Mg, Si, Ca, and Ti. Both the abundances derived from the observed EWs 
and the {\em bona fide} values of \#1657 obtained from the veiled EWs
are in excellent agreement with each other and, most importantly, with the mean ratios of the GC's TO sample (KM11).
The only exception is Si.
Thus, the [$\alpha$/Fe] ratios in \#1657 can be considered fully compatible with NGC~6397's enrichment patterns. 
\begin{center}
\begin{deluxetable}{cccccc}
\tabletypesize{\scriptsize}
\tablecaption{Iron and $\alpha$-element abundance ratios in \#1657}
\tablewidth{0pt}
\tablehead{  \colhead{} & \multicolumn{2}{c}{Value} & \colhead{} & \colhead{} & \colhead{}  \\
\cline{2-3}
 \raisebox{1.5ex}[-1.5ex]{Ratio} & \colhead{(unveiled)} & \colhead{(veiled)} & \raisebox{1.5ex}[-1.5ex]{$\sigma$} &  \raisebox{1.5ex}[-1.5ex]{$N$}&
 \raisebox{1.5ex}[-1.5ex]{$\sigma_{\rm syst}$\tablenotemark{a}}}
\startdata
A(Li)$_{\rm NLTE}$ & \phantom{$-$}4.03 &  \phantom{$-$}4.21 & 0.06 & 2 & 0.14 \\
$[$\ion{Fe}{1}/H] & $-$2.04 & $-$1.93 & 0.27 & 18 & 0.18 \\ 
$[$\ion{Fe}{2}/H] & $-$2.26 & $-$2.09 & \dots & 1 & 0.18 \\ 
$[$\ion{Mg}{1}/\ion{Fe}{1}] & \phantom{$-$}0.22 & \phantom{$-$}0.19 &  0.04 & 2  & 0.10 \\ 
$[$\ion{Si}{1}/\ion{Fe}{1}] & \phantom{$-$}0.34 & \phantom{$-$}0.50 & \dots & 1  & 0.33 \\ 
$[$\ion{Ca}{1}/\ion{Fe}{1}] & \phantom{$-$}0.42 & \phantom{$-$}0.42 &  0.14 & 9  & 0.15 \\ 
$[$\ion{Ti}{1}/\ion{Fe}{1}] & \phantom{$-$}0.15 & \phantom{$-$}0.13 &  0.14 & 4  & 0.21 \\ 
$[$\ion{Ti}{2}/\ion{Fe}{2}] & \phantom{$-$}0.37 & \phantom{$-$}0.37 &  0.22 & 16 & 0.27 
\enddata
\tablenotetext{a}{Based on $\Delta$T$_{\rm eff}$$=\pm250$ K; $\Delta\log g$$=\pm0.2$; 
$\Delta\,\xi$$=\pm0.8$ km\,s$^{-1}$; $\Delta$[M/H]=$\pm$0.2 dex; $\Delta$[$\alpha$/Fe]=0.4 dex.}
\end{deluxetable}
\end{center}

\section{Discussion}
We have discovered a super-Li rich TO star in the metal poor GC NGC~6397 and the origin of its Li abundance, a factor of 100 higher than  the Spite-plateau, remains puzzling. 
In the following we discuss several scenarios that are commonly evoked to explain strong Li-enhancements.

{\em Capture of a substellar body:} 
Accretion of smaller companions, which retained their primordial Li, into the star's convective envelope can elevate the abundance (Ashwell et al. 2005; Pasquini et al. 2007). This would, in turn, yield higher abundances of the refractory elements (Takeda et al. 2001), for which we do not see any evidence in our spectra apart from the marginally higher Fe abundance -- e.g., the line strengths of the Fe peak elements in the TO stars \#1657 and \#13552 are indistinguishable safe for the veiling due to the binary companion. Moreover, the occurrence of planetary systems at this low metallicity is unlikely.  Detection of Be lines in the UV would be crucial to test this scenario. 

{\em Type~II Supernovae (SNe):}  $^7$Li can be produced via the $\nu$-process in $\sim$30 M${\odot}$ core-collapse SNe (e.g., Woosley \& Weaver 1995). 
However, such an event would lead to significantly higher abundances of the hydrostatic elements, while we do not see any evidence for this. The fact that the 
line strengths of the species affected by nucleosynthesis in such massive stars  (e.g., Mg, Ca, Ti) are indistinguishable between the TO stars 
then argues against substantial production of \#1657's Li in the $\nu$-process. 

{\em Diffusion:} 
In the narrow temperature  window of 6900--7100 K the surface convection zone of dwarfs can become enriched by radiative outward acceleration of Li from 
deeper regions (Richard et al. 2005). 
If our T$_{\rm eff}$ was significantly in error and the true temperature of the star was 
within that narrow range, then its  A(Li)$_{\rm NLTE}$ would be even higher, at $\sim$4.40 dex,  and would in fact lie on the Li-peak predicted by 
diffusive models.  As above, the similarity of the Li-rich spectrum  and the other TO stars precludes a severe mis-classification of the spectral type, in particular 
judging by the H$\alpha$ profiles. 
Korn et al. (2007) found evidence of diffusion (via gravitational settling and thermal mixing) in NGC~6397 in terms of temperature-dependent Fe- and 
$\alpha$-abundances. While we note a marginal trend of higher 
[Fe/H] (by $\sim$0.1 dex) and lower [Mg/H] with increasing temperature (KM11), the uncertainties in T$_{\rm eff}$ and 
the abundance ratios of  \#1657 do not allow for a further comparison with diffusion scenarios. 

{\em AGB companion:} 
The CF71 mechanism was originally conceived to result from He-shell flashes that induce convective envelope mixing in AGB stars;  hence Li is freshly produced 
in  the the outer, convected layers. Likewise, early Li-enrichment by the wind of a super-AGB star that experienced 
hot bottom processing can be a possible mechanism (Ventura \& D'Antona 2011).  
If \#1657 is part of a binary system, in which the secondary  has evolved through the AGB phase and transferred material onto the primary, the enhancement of Li should be prominently accompanied by overabundances of $s$-process-elements.  
While we cannot exclude the possibility of a binary nature of this object, we do not find any traces of such patterns: 
the line strengths of prominent features  such as \ion{Ba}{2}~4554\AA~and \ion{Sr}{2}~4077 \AA~are in unison with the remainder of the TO sample,   
thus likely ruling out a contamination of \#1657 with matter processed through AGB-nucleosynthesis. 

{\em Cool bottom processing:} 
Li synthesis through CF71 in a H-shell fusion zone in conjunction with  extra deep envelope mixing in low-mass red giants (i.e., CBP)  is often invoked to explain the nature of  (super-) Li-rich  giants. This  self enrichment process is particularly efficient in metal poor GCs, since their  stars have hotter CNO-burning shells. In our star, mass transfer from  a red giant companion that underwent CBP is an attractive possibility, as this process does not incur enhancements of the $s$-process elements.
 
On the other hand, on the RGB Li is destroyed by proton capture over very short periods, at typical depletion time scales of a few $\times 10^4$ yr (e.g., Kraft et al. 1999; and 
references therein), as the freshly synthesized Li is mixed deeply back into the higher-T regions.
Thus, while the chemical signatures in our star are compatible with an external enrichment by CBP-processed material, an exact and fortunate timing of the accretion process during the common, yet transient phase of Li-enhancement on the RGB must be realized. 

\acknowledgments
We are grateful to A. McWilliam for providing the spectra and to I.B. Thompson for helpful discussions. 
AK thanks the Deutsche Forschungsgemeinschaft for funding from  Emmy-Noether grant  Ko 4161/1.


\begin{thebibliography}{}
%
\bibitem[Abia et al.(1993)]{1993A&A...272..455A} Abia, C., Boffin, H.~M.~J., Isern, J., \& Rebolo, R.\ 1993, \aap, 272, 455 
%
\bibitem[Anderson et al.(2008)]{2008AJ....135.2114A} Anderson, J., et al.\  2008, \aj, 135, 2114 
%
\bibitem[Ashwell et al.(2005)]{2005MNRAS.363L..81A} Ashwell, J.~F., 
Jeffries, R.~D., Smalley, B., Deliyannis, C.~P., Steinhauer, A.,  \& King, J.~R.\ 2005, \mnras, 363, L81 
%
\bibitem[Boesgaard et al.(1998)]{1998ApJ...493..206B} Boesgaard, A.~M., Deliyannis, C.~P., Stephens, A., \& King, J.~R.\ 1998, \apj, 493, 206 
%
\bibitem[Bonifacio et al.(2002)]{2002A&A...390...91B} Bonifacio, P., et al.\ 2002, \aap, 390, 91 
%
\bibitem[Brown et al.(1989)]{1989ApJS...71..293B} Brown, J.~A., Sneden, C., Lambert, D.~L., \& Dutchover, E., Jr.\ 1989, \apjs, 71, 293 
%
\bibitem[Cameron \& Fowler(1971)]{1971ApJ...164..111C} Cameron, A.~G.~W., \& Fowler, W.~A.\ 1971, \apj, 164, 111 
%
\bibitem[Castelli \& Kurucz(2004)]{2004astro.ph..5087C} Castelli, F., \& Kurucz, R.~L.\ 2004 (arXiv:astro-ph/0405087) 
%
\bibitem[Cayrel(1988)]{1988IAUS..132..345C} Cayrel, R.\ 1988, The Impact of Very High S/N Spectroscopy on Stellar Physics, 132, 345 
%
\bibitem[Cyburt et al.(2008)]{2008JCAP...11..012C} Cyburt, R.~H., Fields, B.~D., \& Olive, K.~A.\ 2008, J. Cosmol. Astro-Part. Phys., 11, 12 
%
\bibitem[Deliyannis et al.(2002)]{2002ApJ...577L..39D} Deliyannis, C.~P., Steinhauer, A., \& Jeffries, R.~D.\ 2002, \apjl, 577, L39 
%
\bibitem[Gratton et al.(2004)]{2004ARA&A..42..385G} Gratton, R., Sneden, C., \& Carretta, E.\ 2004, \araa, 42, 385 
%
\bibitem[Gustafsson et al.(2008)]{2008A&A...486..951G} Gustafsson, B., Edvardsson, B., Eriksson, K., J{\o}rgensen, U.~G., Nordlund, {\AA}., \& Plez, B.\ 2008, \aap, 486, 951 
%
\bibitem[Harris(1996)]{1996AJ....112.1487H} Harris, W.~E.\ 1996, \aj, 112, 1487 
%
\bibitem[Hobbs et al.(1999)]{1999ApJ...523..797H} Hobbs, L.~M., Thorburn, J.~A., \& Rebull, L.~M.\ 1999, \apj, 523, 797 
%
\bibitem[Kaluzny(1997)]{1997A&AS..122....1K} Kaluzny, J.\ 1997, \aaps, 122, 1 
%
\bibitem[Koch \& McWilliam(2011)]{2011AJ............} Koch, A., \& McWilliam, A. 2011, AJ, 142, 63 
%
\bibitem[Kochukhov(2008)]{2008A&A...483..557K} Kochukhov, O.\ 2008, \aap, 483, 557 
%
\bibitem[Korn et al.(2007)]{2007ApJ...671..402K} Korn, A.~J., Grundahl, F.,  Richard, O., Mashonkina, L., Barklem, P.~S., Collet, R., Gustafsson, B., \& Piskunov, N.\ 2007, \apj, 671, 402 
%
\bibitem[Kraft et al.(1999)]{1999ApJ...518L..53K} Kraft, R.~P., Peterson, R.~C., Guhathakurta, P., Sneden, C., Fulbright, J.~P., \& Langer, G.~E.\ 1999, \apjl, 518, L53 
%
\bibitem[Lambert \& Reddy(2004)]{2004MNRAS.349..757L} Lambert, D.~L., \& Reddy, B.~E.\ 2004, \mnras, 349, 757 
%
\bibitem[Lind et al.(2009)]{2009A&A...503..545L} Lind, K., Primas, F., Charbonnel, C., Grundahl, F., \& Asplund, M.\ 2009a, \aap, 503, 545 
%
\bibitem[Lind et al.(2009)]{2009A&A...503..541L} Lind, K., Asplund, M., \& Barklem, P.~S.\ 2009b, \aap, 503, 541 
%
\bibitem[Lind et al.(2010)]{2010arXiv1012.0477L} Lind, K., Charbonnel, C.,  Decressin, T., Primas, F., Grundahl, F., \& Asplund, M.\ 2010, arXiv:1012.0477 
%
\bibitem[Monaco et al.(2011)]{2011A&A...529A..90M} Monaco, L., et al.\ 2011, \aap, 529, A90 
%
\bibitem[Palacios et al.(2001)]{2001A&A...375L...9P} Palacios, A., Charbonnel, C., \& Forestini, M.\ 2001, \aap, 375, L9 
%
\bibitem[Pasquini et al.(2007)]{2007A&A...473..979P} Pasquini, L., D{\"o}llinger, M.~P., Weiss, A., Girardi, L., Chavero, C., Hatzes, A.~P., da Silva, L., \& Setiawan, J.\ 2007, \aap, 473, 979 
%
\bibitem[Preston(1994)]{1994AJ....108.2267P} Preston, G.~W.\ 1994, \aj, 108, 2267 
%
\bibitem[Reeves(1970)]{1970Natur.226..727R} Reeves, H.\ 1970, \nat, 226, 727 
%
\bibitem[Rich et al.(2011)]{2011AAS...21715227R} Rich, R.~M., et al.\ 2011, BAAS, 43, \#152.27 
%
\bibitem[Richard et al.(2005)]{2005ApJ...619..538R} Richard, O., Michaud, G., \& Richer, J.\ 2005, \apj, 619, 538 
%
\bibitem[Sackmann \& Boothroyd(1999)]{1999ApJ...510..217S} Sackmann, I.-J., \& Boothroyd, A.~I.\ 1999, \apj, 510, 217 
%
\bibitem[Spite \& Spite(1982)]{1982A&A...115..357S} Spite, F., \& Spite, M.\ 1982, \aap, 115, 357 
%
\bibitem[Smith et al.(1999)]{1999ApJ...516L..73S} Smith, V.~V., Shetrone, M.~D., \& Keane, M.~J.\ 1999, \apjl, 516, L73 
%
\bibitem[Takeda et al.(2001)]{2001PASJ...53.1211T} Takeda, Y., et al.\ 2001, \pasj, 53, 1211 
%
\bibitem[Thompson et al.(2008)]{2008ApJ...677..556T} Thompson, I.~B., et al.\ 2008, \apj, 677, 556 
%
\bibitem[Ventura \& D'Antona(2011)]{2011MNRAS.410.2760V} Ventura, P., \& D'Antona, F.\ 2011, \mnras, 410, 2760 
%
\bibitem[Wasserburg et al.(1995)]{1995ApJ...447L..37W} Wasserburg, G.~J., Boothroyd, A.~I., \& Sackmann, I.-J.\ 1995, \apjl, 447, L37 
%
\bibitem[Woosley \& Weaver(1995)]{1995ApJS..101..181W} Woosley, S.~E., \& Weaver, T.~A.\ 1995, \apjs, 101, 181 
%
\end{thebibliography}
\end{document}